\documentclass[11pt,a4paper]{article}
\pdfoutput=1
\usepackage{amsmath,amssymb,tikz}
\usepackage{graphicx}
 \allowdisplaybreaks
\def\be{\begin{equation}}
\def\ee{\end{equation}}
\def\ba#1\ea{\begin{align}#1\end{align}}
\def\bg#1\eg{\begin{gather}#1\end{gather}}
\def\bm#1\em{\begin{multline}#1\end{multline}}
\def\bmd#1\emd{\begin{multlined}#1\end{multlined}}

\def\({\left(}
\def\){\right)}
\def\[{\left[}
\def\]{\right]}

\def \be {\begin{equation}}
\def \ee {\end{equation}}
\def \ba {\begin{array}}
\def \ea {\end{array}}
\def \bea{\begin{eqnarray}}
\def \eea{\end{eqnarray}}

\def\bea{\begin{eqnarray}}
\def\eea{\end{eqnarray}}

\newcommand{\bit}{\begin{itemize}}  \newcommand{\eit}{\end{itemize}}
\newcommand{\ben}{\begin{enumerate}}  \newcommand{\een}{\end{enumerate}}

\long\def\symbolfootnote[#1]#2{\begingroup%
\def\thefootnote{\fnsymbol{footnote}}\footnote[#1]{#2}\endgroup}
\newcommand{\sysu}{{\it School of Physics and Astronomy, Sun Yat-Sen University, 2 Daxue Road, Zhuhai 519082, China}}
\begin{document}
%%%%%%%%%%%%%%%%%%%%%%%%%%%%%%%%%%%%%%%%%%%
\thispagestyle{empty}
%%%%%%%%%%%%%%%%%%%%%%%%%%%%%%%%%%%%%%%%%%%
%%%%%%%%%%%%%%%%%%%%%%%%%%%%%%%%%%%%%%%%%%%
%\begin{flushright}
%\hfill{AEI-2015-xxx}
%\hfill{ NCTS-TH/1702}
%\end{flushright}
%%%%%%%%%%%%%%%%%%%%%%%%%%%%%%%%%%%%%%%%%%%
\begin{center}

~\vspace{20pt}

{\Large\bf Holographic Anomalous Current at a Finite Temperature}

\vspace{25pt}

Jian-Guo Liu, Rong-Xin Miao ${}$\symbolfootnote[1]{Email:~\sf
  miaorx@mail.sysu.edu.cn}

\vspace{10pt}${}$\sysu

\vspace{2cm}

\begin{abstract}
Weyl anomaly leads to novel anomalous currents in a spacetime with boundaries. Recently it is found that the anomalous current can be significantly enhanced by the high temperature for free theories, 
which could make the experimental measurement easier.
 In this paper, we investigate holographic anomalous currents at a finite temperature. It is found that the holographic current is still enhanced by the high temperature in dimensions higher than three. However, the temperature dependence is quite different from that of free theories. This may be due to the fact that the holographic CFT is strongly coupled and there is non-zero resistance in the holographic model. Remarkably, the temperature dependence of holographic anomalous currents is universal in the high temperature limit, which is independent of the choices of background magnetic fields. 
\end{abstract}

\end{center}

%%%%%%%%%%%%%%%%%%%%%%%%%%%%%%%
\newpage
\setcounter{footnote}{0}
\setcounter{page}{1}
%%%%%%%%%%%%%%%%%%%%%%%%%%%%%%%

\tableofcontents
%%%%%%%%%%%%%%%%%%%%%%%%%%%%%%%

\section{Introduction}

Due to Weyl anomaly \cite{Duff:1993wm}, an external electromagnetic field can induce novel anomalous currents in a conformally flat space \cite{Chernodub:2016lbo, Chernodub:2017jcp} and a spacetime with boundaries \cite{Chu:2018ksb,Chu:2018ntx}. Similar to Casimir effect \cite{Casimir:1948dh,Plunien:1986ca,Bordag:2001qi}, the anomalous currents arise from the effect of the background gravitational field and the boundary on the quantum fluctuations of the vacuum. Other anomaly-induced transports \cite{review} include chiral magnetic effect (CME) \cite{Vilenkin:1995um,
Vilenkin:1980fu, Giovannini:1997eg, alekseev, Fukushima:2012vr} and chiral vortical effect (CVE) \cite{Kharzeev:2007tn,Erdmenger:2008rm,
 Banerjee:2008th,Son:2009tf,Landsteiner:2011cp,Golkar:2012kb,Jensen:2012kj}. See also \cite{Chu:2018fpx,Chu:2019rod,Miao:2017aba,Miao:2018dvm,Chernodub:2018ihb,Chernodub:2019blw,Ambrus:2019khr,Zheng:2019xeu,Miao:2018qkc,Chu:2020mwx,Chu:2020gwq,Hu:2020puq,Kawaguchi:2020kce,Kurkov:2020jet,Kurkov:2018pjw,Fialkovsky:2019rum,Vassilevich:2003xt,McAvity:1990we,JohnP,FalKovskii} for related works. 

In this paper, we focus on the anomalous current in the spacetime with a boundary \cite{Chu:2018ksb,Chu:2018ntx}. In four dimensions, it takes a universal form
\begin{eqnarray}\label{typeIIcurrent}
<J^{\mu}>=\frac{-2\beta F^{\mu\nu}n_{\nu}}{x} +..., \ x\sim 0,
\end{eqnarray}
near the boundary. Here $\beta$ is the beta function, $F^{\mu\nu}$ are the field strength, $x$ is the proper distance to the boundary, $n_{\mu}$ are the normal vectors and $...$ denotes higher order terms in $O(x)$.  Note that (\ref{typeIIcurrent}) applies to not only the conformal field theory (CFT) but also the general quantum field theory (QFT). That is because it is derived from Weyl anomaly \cite{Chu:2018ksb}, which is well-defined for the general QFT \cite{Duff:1993wm,Brown:1976wc,Casarin:2018odz}.  

Unfortunately, the anomalous current is heavily suppressed by the mass and the distance to the boundary  \cite{Chernodub:2018ihb,Hu:2020puq}, which makes it difficult to be measured in laboratory. Remarkably, recently it is found that the anomalous current can be greatly enhanced by the high temperature for free theories \cite{Guo:2021dzz}, which could make the experimental measurement easier. For free theories, the anomalous current is proportional to the temperature in the high temperature limit
\begin{eqnarray}\label{freetheory}
\lim_{T\to \infty}<J>\ \sim \ T.
\end{eqnarray}
This means, for any given charge carrier with fixed mass, one can always produce a detectable anomalous current by increasing the temperature. 

In this paper, we investigate the holographic anomalous current at a finite temperature. We find that the holographic current is still enhanced by the high temperature in dimensions higher than three. However, the temperature dependence is quite different from that of free theories. See (\ref{currentsect1},\ref{holocurrentsect2},\ref{currentsect3}) for example. The holographic current decreases with the temperature in three dimensions, while it increases with the temperature in dimensions higher than four. In four dimensions, the absolute value of current first decreases and then increases with the temperature. In the high temperature limit, the holographic anomalous current takes a universal form, 
 \begin{eqnarray}\label{currentsect1}
\lim_{T\to \infty}<J>\ \sim  \begin{cases}
 \ \ T^{d-4} ,\ \ \ \ \ \ \ \ \ \ \ d\ne 4,\\
\ \  \log ( T),\ \ \ \ \ \ \ \ \ d=4,
\end{cases}
\end{eqnarray}
which is independent of the choices of external magnetic fields. 

The paper is organized as follows.
In section 2, by choosing a suitable external magnetic field, we derive an exact expression of the holographic anomalous current. We find that the holographic anomalous current is enhanced by the high temperature in dimensions higher than three. In section 3, we discuss the general background magnetic fields and derive the holographic anomalous current in the high temperature limit. In section 4, we present some numerical results. Finally, we conclude with some open questions in section 5.

\section{Holographic current I: exact result}

The main purpose of this paper is to explore the temperature dependence of holographic anomalous currents \cite{Chu:2018ntx,Miao:2018qkc}.
 To warm up, let us first study a special case that the background magnetic field is given by $B=2 b_a x$, where $b_a$ are a constant and $x$ is the distance to the boundary.
 As it is shown below, we can derive exact expressions of anomalous currents for this case. 

\begin{figure}[t]
\centering
\includegraphics[width=12cm]{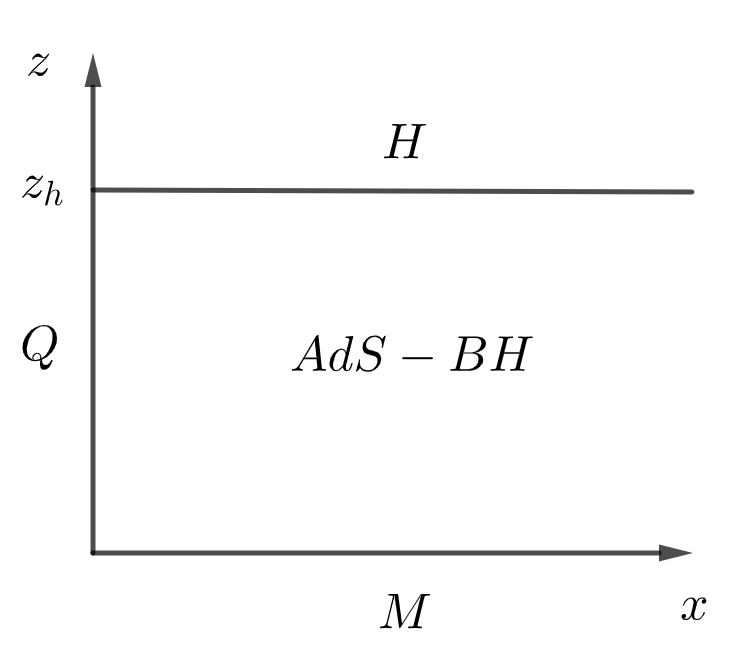}
\caption{Geometry of the holographic model. The bulk is an AdS-Schwarzschild black hole,  the bulk boundary $Q$, the AdS boundary $M$ and the horizon $H$ are located at $x=0$, $z=0$ and $z=z_h$, respectively.}
\label{FigureAdSBCFT}
\end{figure}

For simplicity, we focus on the probe limit, where the background spacetime is given by AdS-Schwarzschild black hole,
\begin{eqnarray}\label{BHmetric}
ds^2=\frac{\frac{dz^2}{1-\frac{z^d}{z_h^d}}-(1-\frac{z^d}{z_h^d})dt^2+dx^2+dy_a^2}{z^2}.
\end{eqnarray}
Here the bulk boundary $Q$, the AdS boundary $M$ and the horizon $H$ are located at $x=0$, $z=0$ and $z=z_h$, respectively.  See Fig. \ref{FigureAdSBCFT} for the geometry of our holographic model. See also \cite{Takayanagi:2011zk} for more explanations of the geometry in AdS/BCFT. 

The temperature of the black hole (\ref{BHmetric}) is 
\begin{eqnarray}\label{tem}
T=\frac{d}{4 \pi  z_h}.
\end{eqnarray}
For simplicity, we choose the following gauges for the bulk Maxwell's fields \cite{Chu:2018ntx}
\begin{eqnarray}\label{bulkvector}
A_z=A_t=A_x=0,\ \ A_a= A_a(z,x).
\end{eqnarray}
Now the bulk Maxwell's equations become
\begin{eqnarray}\label{bulkvectorEOM}
\left(3 Z^d+d-3\right) A_a{}^{(1,0)}(Z,x)+Z \left(Z^d-1\right) A_a{}^{(2,0)}(Z,x)-Z z_h^2 A_a{}^{(0,2)}(Z,x)=0,
\end{eqnarray}
where we have made the coordinate transformation $Z=z/z_h$.  Following \cite{Takayanagi:2011zk}, we impose Neumann boundary condition (NBC) on the bulk boundary $Q$
\begin{eqnarray}\label{NBC}
\partial_x A_a(Z,x)|_{x=0} =0,
\end{eqnarray}
which is equivalent to the absolute BC, i.e.,  $F_{xa}|_{x=0}=0$.

We take the following ansatz of $A_a$, 
\begin{eqnarray}\label{ansatzAa}
A_a(Z,x) = b_a x^2+d_a+ \bar{A}_a(Z),
\end{eqnarray}
which automatically obeys the NBC (\ref{NBC}).
Here $b_a, d_a$ are constants and the background magnetic field on the AdS boundary is 
\begin{eqnarray}\label{B}
B=F_{xa}|_{Z\to 0}=2b_a x.
\end{eqnarray}
Substituting (\ref{ansatzAa}) into (\ref{bulkvectorEOM}), we get
\begin{eqnarray}\label{EOMofAa}
Z \left(Z^d-1\right) \bar{A}_a''(Z)+\left(3 Z^d+d-3\right) \bar{A}_a'(Z)-2 b_a Z  z_h^2=0,
\end{eqnarray}
which can be solved as
\begin{eqnarray}\label{SolutionAa}
\bar{A}_a=\begin{cases}
c_{2a}+\frac{c_{1a} Z^{d-2} \, _2F_1\left(1,\frac{d-2}{d};2-\frac{2}{d};Z^d\right)}{d-2}+\frac{b_a Z^2 \text{zh}^2 \, _2F_1\left(1,\frac{2}{d};\frac{d+2}{d};Z^d\right)}{d-4},\  \ \ \ \ \ \ \ \ \ \ \ \ \ \ \ \ \ \ \ \ \ \ \ \ \ d\ne 4\\
c_{2a}+\frac{c_{1a}}{4}  \log \left(\frac{Z^2+1}{1-Z^2}\right)-\frac{b_a z_h^2 }{8} \left(\text{Li}_2\left(Z^4\right)-4 \text{Li}_2\left(Z^2\right)+4 \log (Z) \log \left(\frac{Z^2+1}{1-Z^2}\right)\right), \ d=4
\end{cases}
\end{eqnarray}
where $ \, _2F_1$ is the hypergeometric function, $\text{Li}_2$ is the polylogarithm function and $c_{1a}, c_{2a}$ are integral constants. Note that $c_{2a}$ can be absorbed into the definition of $d_a$ (\ref{ansatzAa}). Thus we can set $c_{2a}=0$ without loss of generality. We impose the natural boundary condition on the horizon 
\begin{eqnarray}\label{naturlaBC}
A_a(Z, x)|_{Z=1} \ \text{is finite},
\end{eqnarray}
which fixes the left integral constant as
\begin{eqnarray}\label{integralconstant}
c_{1a}=\begin{cases}
\frac{2 b_a z_h^2}{4-d},\ \ \ d\ne 4,\\
0,\ \ \ \ \ \ \ \ d=4.
\end{cases}
\end{eqnarray}

The holographic current can be read off from the asymptotic solutions of bulk vectors near the AdS boundary $z=0$
\begin{eqnarray}\label{holocurrentformula}
A_a= A_a(z=0) + ... + z^{d-2} \left(\frac{J_a}{d-2} + f_a \ln z \right)+...,
\end{eqnarray}
where $J_a$ is the holographic current and $f_a$ is an irrelevant function which appears only in even dimensions. Expanding the solutions (\ref{SolutionAa}) near $z=0$ and comparing with (\ref{holocurrentformula}), 
we finally obtain the holographic anomalous current
\begin{eqnarray}\label{holocurrentsect2}
J_a=\begin{cases}
\frac{ 2^{2 d-7} \pi ^{d-4}}{(4-d) d^{d-4}} b_a \ T^{d-4},\ \ \ \ \ \ \ \ d\ne 4,\\
b_a \big(1-2 \log (\pi  T)\big),\ \ \ \ \ \ \ \ d=4.
\end{cases}
\end{eqnarray}
In dimensions higher than four, the holographic current increases with the temperature, while in dimensions lower than four, the anomalous current decreases with the temperature. In four dimensions, the absolute value of current first decreases and then increases with the temperature.

\section{Holographic current II: perturbative result}

In the above section, we focus on a special kind of external magnetic field (\ref{B}) and derive an exact expression of the holographic anomalous current. In this section, we discuss the currents induced by general background magnetic fields.  In the high temperature limit, we find that the temperature dependence of holographic currents is universal and is still given by (\ref{holocurrentsect2}).  

The general bulk Maxwell's fields obeying NBC (\ref{NBC}) take the following form
\begin{eqnarray}\label{ansatzAasect3}
A_a(Z,x) = \int dk F(k)\cos(k x) \hat{A}_a(Z),
\end{eqnarray}
where $F(k)$ is an arbitrary function as long as it defines a convergent integral.  From (\ref{ansatzAasect3}), we read off the background magnetic fields on the AdS boundary
\begin{eqnarray}\label{Bsect3}
B=-\int dk k F(k)\sin(k x) \hat{A}_a(0).
\end{eqnarray}
Substituting (\ref{ansatzAasect3}) into (\ref{bulkvectorEOM}), we obtain
\begin{eqnarray}\label{EOMsect3}
Z \left(Z^d-1\right) \hat{A}_a''(Z)+\left(3 Z^d+d-3\right) \hat{A}_a'(Z)+k^2 Z z_h^2 \hat{A}_a(Z)=0.
\end{eqnarray}
Note that the above equation is quite similar to (\ref{EOMofAa}), only the last term is different.  Unfortunately, unlike  (\ref{EOMofAa}), the above equation cannot be solved analytically.  For simplicity, we investigate the perturbation solutions in this section, 
and leave the discussions of numerical solutions to next section.

 In the high temperature limit, $z_h=d/(4\pi T)$ is a small parameter. Thus we can expand the bulk vectors $ \hat{A}_a(Z)$ in powers of $z_h^2$,
\begin{eqnarray}\label{Aapersect3}
\hat{A}_a(Z)= \hat{A}_a^{(0)}(Z)+ z_h^2  \hat{A}_a^{(1)}(Z) +O(z_h^4).
\end{eqnarray}
Let us solve (\ref{EOMsect3}) order by order in $O(z_h^2)$.  One can check that $\hat{A}_a^{(0)}(Z)$ must be a constant
\begin{eqnarray}\label{Aapersect3A0}
\hat{A}_a^{(0)}(Z)=c_a
\end{eqnarray}
 in order to satisfy the EOM (\ref{EOMsect3}) and the natural boundary condition (\ref{naturlaBC}) at the same time. Substituting (\ref{Aapersect3},\ref{Aapersect3A0}) into (\ref{EOMsect3}), we get the EOM of $\hat{A}_a^{(1)}(Z)$, which is exactly the same as (\ref{EOMofAa}) provided that we identify $-2b_a z_h^2$ with $c_a k^2$.  Now following the approach of sect. 2, we can derive the holographic current as 
 \begin{eqnarray}\label{currentsect3}
J_a= -\frac{c_a+ O(1/T^2)}{2}\int dk F(k) k^2\cos(k x) \begin{cases}
\frac{ 2^{2 d-7} \pi ^{d-4}}{(4-d) d^{d-4}}  \ T^{d-4} ,\ \ \ \ \ \ \ \ d\ne 4,\\
 \big(1-2 \log (\pi  T)\big),\ \ \ \ \ \ \ \ d=4.
\end{cases}
\end{eqnarray}
It is remarkable that the temperature dependence of the holographic current is universal, which is independent of the choices of background vector fields in the high temperature limit. 

\section{Holographic current III: numerical result}

 In this section, we discuss holographic anomalous currents at general temperatures. The main task is to solve (\ref{EOMsect3}) numerically. To do so, we need to specify the boundary condition near the horizon $Z=1$.  Assume that  $\hat{A}_a(Z)$ takes the following form near the horizon
 \begin{eqnarray}\label{Aasect4}
\hat{A}_a=a_{0a}+ a_{1a} (1-Z)+ a_{2a} (1-Z)^2+ a_{3a} (1-Z)^3+O(1-Z)^4,
\end{eqnarray}
 where $a_{i a}$ are constants to be determined.  Imposing the natural boundary condition (\ref{naturlaBC}), we solve
 \begin{eqnarray}\label{aisect4}
&&a_{1a}=\frac{ k^2 z_h^2}{d}  a_{0a},\nonumber\\
&&a_{2a}=\frac{\left(2 d k^2 z_h^2+k^4 z_h^4\right)}{4 d^2} a_{0a},\nonumber\\
&&a_{3a}=\frac{k^2 z_h^2 \left(-4 (d-5) d^2+d (d+5) k^2 z_h^2+k^4 z_h^4\right)}{36 d^3} a_{0a},
\end{eqnarray}
where $a_{0a}$ is a free parameter, which can be set to be $a_{0a}=1$.  Now the boundary conditions of the differential equation (\ref{EOMsect3}) become
 \begin{eqnarray}\label{BCsect4}
&&\hat{A}_a (1-\epsilon)= a_{0a}+ a_{1a} \epsilon+ a_{2a} \epsilon^2+ a_{3a} \epsilon^3+O(\epsilon^4),\nonumber\\
&&\hat{A}'_a (1-\epsilon)=-a_{1a} -2a_{2a} \epsilon -3a_{3a} \epsilon^2+O(\epsilon^3),
\end{eqnarray}
where $\epsilon$ is a small constant chosen for the numerical calculations, and $a_{ia}$ are given by (\ref{aisect4}) with $a_{0a}=1$. 

\begin{figure}[t]
\centering
\includegraphics[width=6cm]{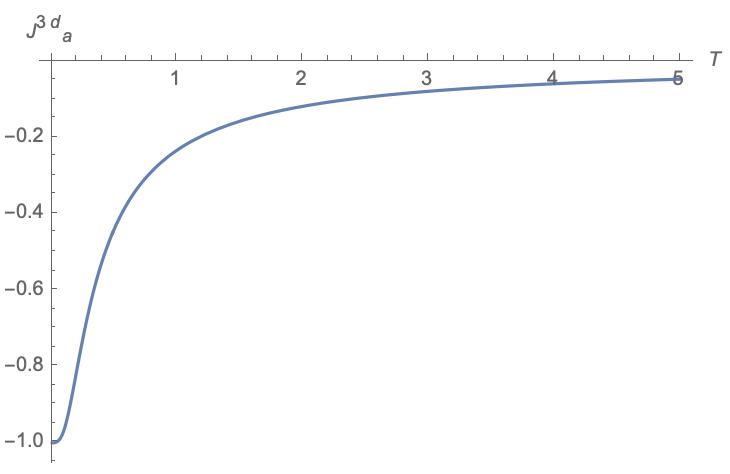}
\includegraphics[width=8cm]{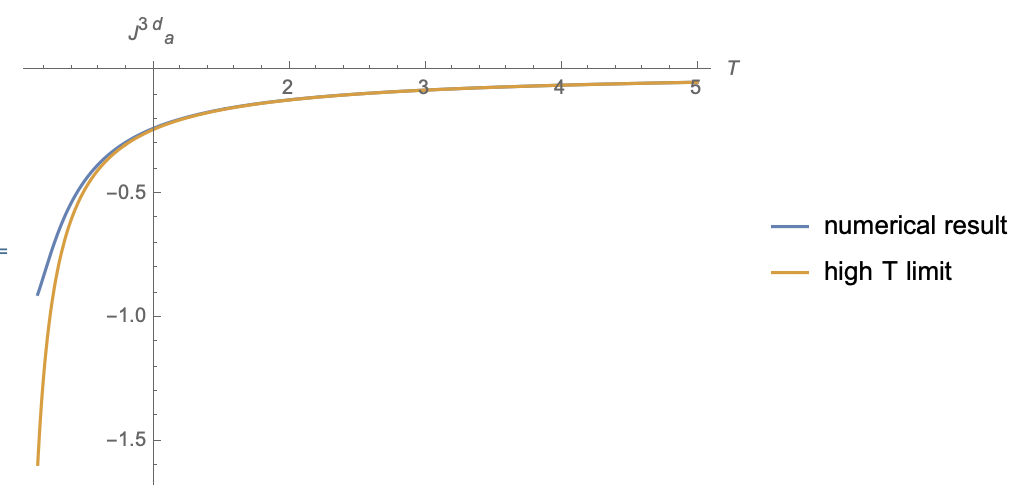}
\caption{Holographic anomalous current at finite temperature in three dimensions (Left); Compare the numerical result  with that of high temperature limit $-\frac{3 k^2}{4 \pi  T}$ (Right). The holographic current decreases with temperature in three dimensions. }
\label{Figure3dCurrent}
\end{figure}

\begin{figure}[t]
\centering
\includegraphics[width=6cm]{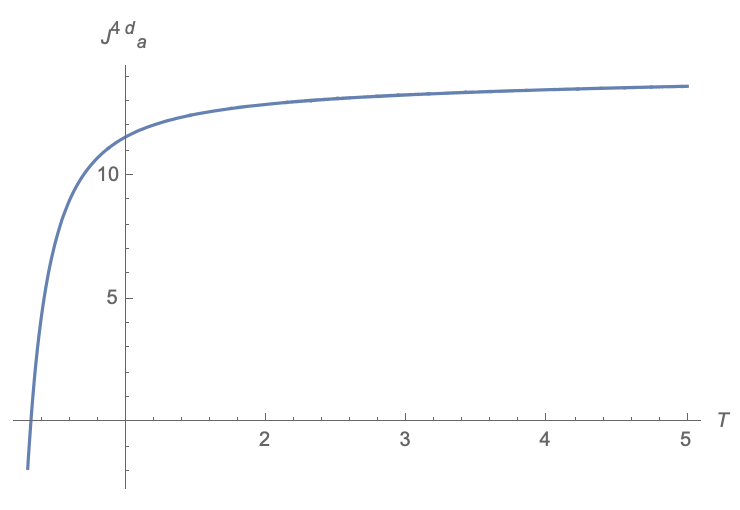}
\includegraphics[width=9cm]{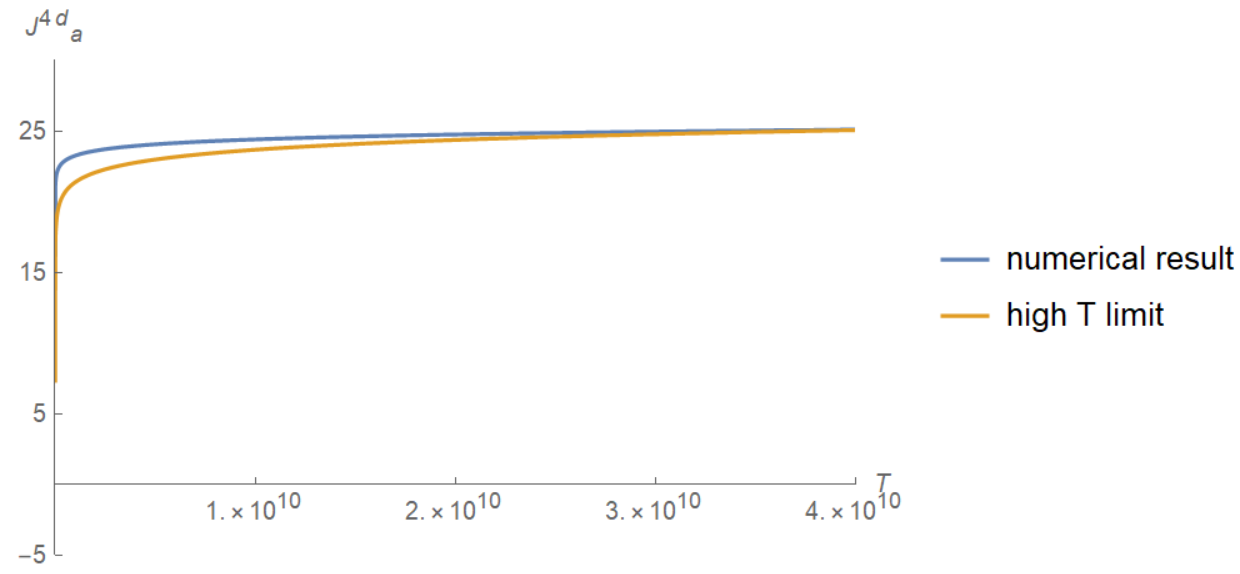}
\caption{Holographic anomalous current at finite temperature in four dimensions (Left); Compare the numerical result  with that of high temperature limit $k^2 \left(\log (\pi  T)-\frac{1}{2}\right)$ (Right). The absolute value of current first decreases and then increases with the temperature in four dimensions. }
\label{Figure4dCurrent}
\end{figure}

\begin{figure}[t]
\centering
\includegraphics[width=6cm]{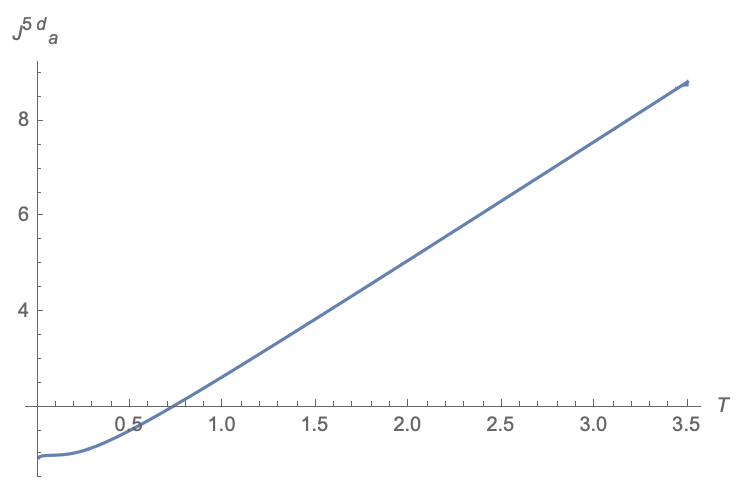}
\includegraphics[width=8cm]{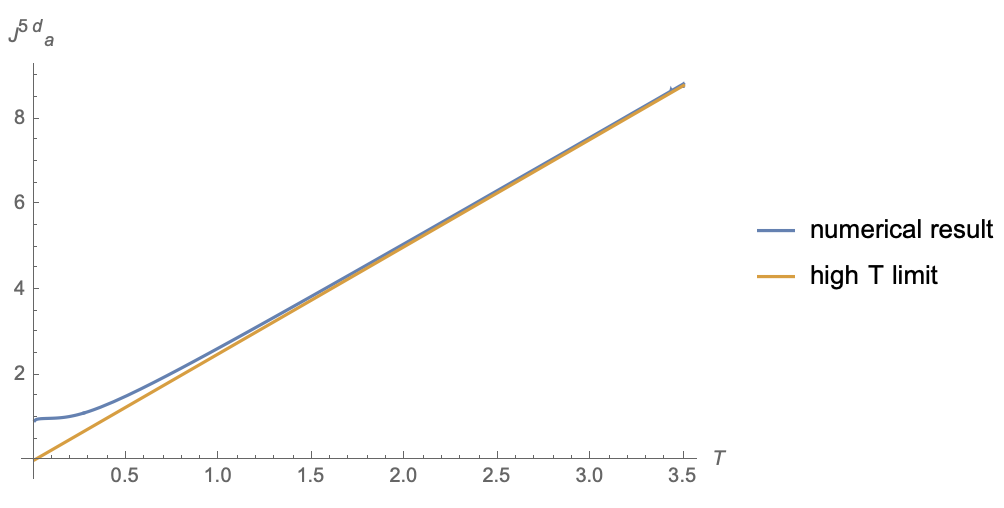}
\caption{Holographic anomalous current at finite temperature in five dimensions (Left); Compare the numerical result  with that of high temperature limit $\frac{4}{5} \pi  k^2 T$ (Right). The holographic current increases with temperature in five dimensions. }
\label{Figure5dCurrent}
\end{figure}

Now it is straightforward to numerically solve the differential equation (\ref{EOMsect3}) together with the BCs (\ref{BCsect4}). Once we obtain the solutions, we can derive the holographic anomalous current from the asymptotic solutions (\ref{holocurrentformula}) near the AdS boundary. Let us draw some figures to show the temperature dependence of the holographic currents.  Without loss of generality, we set $k=a_{0a}=1$ for all the figures. See Fig. \ref{Figure3dCurrent}, Fig. \ref{Figure4dCurrent}, Fig. \ref{Figure5dCurrent} for examples. It is found that the holographic anomalous currents in these figures match exactly the results (\ref{currentsect3}) of sect.3 in the high temperature limit. 
This can be regarded as a double check of our results. Besides, the current decreases with the temperature in three dimensions, while it increases with the temperature in dimensions higher than four. In four dimensions, the absolute value of current first decreases and then increases with the temperature.

\section{Conclusions and Discussions}
In this paper, we investigate the holographic anomalous current at a finite temperature. For the external magnetic field $B\sim x$, we derive an exact expression of the holographic current. As for general background magnetic fields, we obtain perturbative and numerical results. It is remarkable that the temperature dependence of the holographic anomalous current is universal in the high temperature limit, which is independent of the choices of background magnetic fields. Similar to the case of free theories, the holographic anomalous current is still enhanced by the high temperature in dimensions higher than three. However, the temperature dependence  is quite different from that of free theories. The reasons may lie in the fact that the holographic CFT is strongly coupled and there is non-zero resistance in the holographic model \cite{Hartnoll:2009sz}. In this paper, we focus on the probe limit, where the background spacetime is given by the AdS-Schwarzschild black hole. 
It is interesting to study the back-reactions and the charged black holes to see if the universal temperature dependence of holographic currents still holds or not. It is also interesting to investigate the low temperature regions carefully. Finally, the temperature dependence of the anomalous Fermi condensation \cite{Chu:2020mwx,Chu:2020gwq} is also a problem worth exploring. We leave these problems to future work.

\section*{Acknowledgements}
We thank C. S. Chu and M. X. Liu for helpful discussions and comments. 
R. X. Miao acknowledges the supports from National Natural Science Foundation of China (No. 11905297) and Guangdong Basic and Applied Basic Research Foundation (No.2020A1515010900).

\appendix

\end{document}